\documentclass[12pt,preprint]{aastex}

\def\plotfiddle#1#2#3#4#5#6#7{\centering \leavevmode
    \vbox to#2{\rule{0pt}{#2}}
    \includegraphics{#1}}

\newcommand{\Msun}{M_{\odot}}

\newcommand{\Msunperyr}{M_{\odot}\,{\rm yr}^{-1}}

\newcommand{\persqcm}{\rm \,cm^{-2}}

\newcommand{\kms}{\,{\rm km}\,{\rm s}^{-1}}

\def\micron{\hbox{$\,\mu$m}}

\shorttitle{[NeII] Emission from AA Tau and GM Aur}
\shortauthors{Najita et al.}

\begin{document}


\title{High Resolution Spectroscopy of [NeII] Emission from AA Tau
and GM Aur\footnote{Based on observations obtained at the Gemini
Observatory, which is operated by the Association of Universities
for Research in Astronomy, Inc., under a cooperative agreement with
the NSF on behalf of the Gemini partnership: the National Science
Foundation (United States), the Science and Technology Facilities
Council (United Kingdom), the National Research Council (Canada),
CONICYT (Chile), the Australian Research Council (Australia),
Ministrio da Cincia e Tecnologia (Brazil) and SECYT (Argentina).}}


\author{Joan R. Najita, Greg W. Doppmann}
\affil{National Optical Astronomy Observatory, 950 N. Cherry Ave., 
Tucson, AZ 85719} 

\author{Martin A. Bitner}
\affil{Space Telescope Science Institute, 3700 San Martin Dr., Baltimore, MD 21218}

\author{Matthew J. Richter} 
\affil{Physics Department, University of California at Davis, Davis, CA 95616}

\author{John H. Lacy, Daniel T. Jaffe}
\affil{Department of Astronomy, University of Texas at Austin, Austin, TX 78712}

\author{John S. Carr}
\affil{Naval Research Laboratory, Code 7213, Washington, DC 20375}

\author{Rowin Meijerink, Geoffrey A. Blake}
\affil{Division of Geological and Planetary Sciences, 
California Institute of Technology, MS 150-21, Pasadena, CA 91125}

\author{Gregory J. Herczeg}
\affil{Max-Planck-Institut f\"ur extraterrestriche Physik, 
Postfach 1312, D-85741 Garching, Germany}

\and

\author{Alfred E. Glassgold}
\affil{Astronomy Department, University of California, Berkeley, CA 94720}


%\altaffiltext{1}{Visiting Astronomer, Cerro Tololo Inter-American Observatory.
%CTIO is operated by AURA, Inc.\ under contract to the National Science
%\altaffiltext{2}{Society of Fellows, Harvard University.}
%\altaffiltext{3}{present address: Center for Astrophysics,
%    60 Garden Street, Cambridge, MA 02138}
%\altaffiltext{4}{Visiting Programmer, Space Telescope Science Institute}
%\altaffiltext{5}{Patron, Alonso's Bar and Grill}

\begin{abstract}
We present high resolution ($R$=80,000) spectroscopy of [NeII] 
emission from two young stars, GM Aur and AA Tau, which have 
moderate to high inclinations. 
The emission from both sources appears centered near the 
stellar velocity and is broader than the [NeII] emission measured 
previously for the face-on disk system TW Hya.  These properties 
are consistent with a disk origin for the [NeII] emission 
we detect, 
with disk rotation (rather than photoevaporation or 
turbulence in a hot disk atmosphere) playing the dominant role 
in the origin of the line width. 
In the non-face-on systems, the [NeII] emission is narrower than 
the CO fundamental emission from the same sources.  If the widths of both 
diagnostics are dominated by Keplerian rotation, this suggests that 
the [NeII] emission arises from larger disk radii on average than 
does the CO emission.  
The equivalent width of the [NeII] emission we detect is less 
than that of the spectrally unresolved [NeII] feature in the 
{\it Spitzer} spectra of the same sources.  Variability 
in the [NeII] emission or the mid-infrared continuum, 
a spatially extended [NeII] component, or a very (spectrally) broad 
[NeII] component might account for the difference in the 
equivalent widths. 
\end{abstract}





\section{Introduction}

The [NeII] $12.8\micron$ emission line has been suggested 
as a potential new probe of the planet formation region of 
circumstellar disks.  It is a potentially powerful diagnostic 
because the neon in disks is expected to be fully in the gas phase
and in atomic form.  In addition, the [NeII] $12.8\micron$ 
line probes warm, ionized gas, conditions which are believed 
to characterize the upper atmosphere 
of the inner disks surrounding classical T Tauri stars 
(Glassgold, Najita, \& Igea 2007; Meijerink, Glassgold, \& Najita 2008) 
and disk photoevaporative flows (Alexander 2008). 
Because it is sensitive to low column densities of gas, [NeII] 
may also be a useful probe of residual gas surrounding 
weak-line T Tauri stars or gas in the optically thin regions of 
transitional disks.  
Glassgold et al.\ (2007) predicted that the inner regions 
($< 20$\,AU) of classical T Tauri disks that are irradiated by stellar 
X-rays would produce strong [NeII] emission that could be detected 
with the {\it Spitzer Space Telescope}. 
Comparably strong [NeII] emission has indeed been detected, 
apparently commonly, in {\it Spitzer} spectroscopy of T Tauri stars 
(Pascucci et al.\ 2007; 
Lahuis et al.\ 2007;  
Ratzka et al.\ 2007; 
Espaillat et al.\ 2007).

While the rough agreement between the predicted and observed 
[NeII] line strengths supports the interpretation of a disk 
origin for the emission, stronger confirmation can be obtained 
from a study of resolved [NeII] line profiles. 
Herczeg et al.\ (2007) previously reported high 
resolution spectroscopy of [NeII] emission from TW Hya. 
TW Hya is a face-on disk system, 
with an inclination of $i=4$ degrees for the inner disk 
(Pontoppidan et al.\ 2008; see also Qi et al.\ 2004).  
It is also a ``transition object'', a T Tauri star whose 
spectral energy distribution (SED) 
indicates that the disk continuum is optically thin within a 
given radius.  In the case of TW Hya, modeling of the SED indicates 
that the disk is optically thin in the continuum 
($\lambda > 1\micron$)
within $\sim 4$\,AU (Calvet et al.\ 2002; cf. Ratzka et al.\ 2007). 

In their study of the [NeII] emission from TW Hya, 
Herczeg et al.\ (2007) found that the emission was centered at 
the stellar velocity, consistent with a disk origin. 
However, the emission was also significantly broader 
(FWHM $\sim 21 \kms$) than has been predicted theoretically 
for a near face-on disk 
(FWHM $< 5\kms$; Meijerink et al.\ 2008; Glassgold et al. 2007).
Herczeg et al.\ suggested that the observed line width 
could be explained if the emission arises from 
(1) a rotating disk at much smaller disk radii ($\sim 0.1$\,AU)
than in the model; 
(2) a disk with a significant turbulent velocity component; or 
(3) a face-on disk undergoing photoevaporation.  
While each of these scenarios might account for the observed 
emission from a face-on system such as TW Hya, Herczeg et al.\ (2007) 
proposed that these scenarios could be tested by observing 
[NeII] emission from higher inclination disk systems. 

A more recent paper (van Boekel et al.\ 2009) illustrates 
another possible origin for the [NeII] emission from T Tauri stars. 
High resolution spectroscopy of the [NeII] 
emission from the T Tau triplet reveals that the emission 
is spatially extended and associated with a known outflow 
in the system.  This suggests that [NeII] emission may arise primarily 
in outflows rather than disks in systems with strong outflow 
activity.

To investigate the scenarios proposed by Herczeg et al.\ (2007) 
for the origin of [NeII] emission in T Tauri stars,  
we present here high resolution spectroscopy of the [NeII] 
emission from AA Tau and GM Aur, 
T Tauri stars that are not known to show strong outflow activity. 
Apart from its high inclination ($i = 75$ degrees; Bouvier et al.\ 1999), 
AA Tau is a typical classical T Tauri star. 
GM Aur is an actively accreting 
transition object that is viewed at an intermediate 
inclination ($i=54$ degrees; Simon et al.\ 2000).
The SED of GM Aur indicates that the disk is optically thin 
in the mid-infrared continuum, suggesting that 
the disk is devoid of small dust grains in the radial range 
5--24\,AU (Calvet et al.\ 2005).

\section{Observations and Data Reduction}

The observations were carried out using TEXES in its high-resolution
mode ($R=80,000$) with a 0.6\arcsec\ wide slit 
on the Gemini North telescope on 29-30 October 2007 
under program ID GN-2007B-C-5. 
The background sky emission was removed by nodding the source along 
the slit and subtracting adjacent nod positions.  
We corrected for telluric absorption with observations of 
$\alpha$ CMa obtained at a similar airmass 
as that of the science targets  
(the difference in airmass was $< 0.2$). 
The spectra were flux calibrated using the low resolution 
{\it Spitzer Space Telescope} spectra reported by 
Furlan et al.\ (2006).
Roughly every 10 minutes, we took
a series of calibration frames including blank sky and an ambient
temperature blackbody.  The blackbody observations were used for
flatfielding the data, while the sky emission line observations were
used for wavelength calibration.  The standard TEXES pipeline
(Lacy et al.\ 2002) which produces wavelength-calibrated one dimensional
spectra, was used to reduce the data.
The wavelength solutions have an accuracy in velocity of $\sim 1\kms$.  

One challenge associated with these observations is that 
the [NeII] line at 12.81 $\mu$m falls in a gap between two TEXES
spectral orders.  By tilting a mirror within the instrument, we are
able to cover the gap by shifting the spectrum on the detector and 
thereby accessing the [NeII] line at the long wavelength end of 
one order or the short wavelength end of the next order.  
In the first case, the blaze efficiency rises toward the blue 
side of the line, while in the second case, it rises toward the 
red side of the line. 

During the night of 29 October, we shifted the
optics to enhance the sensitivity on the red side of 
the [NeII]
line.  We spent 2590 seconds of on-source integration time on GM
Aur and 3238 seconds on AA Tau in this instrument setup.  The
observed GM Aur line appeared narrow, symmetric, and centered at
the stellar velocity while the emission from AA Tau was broad and
redward of the stellar velocity.  This led us to search for a
line profile component that might be present blueward of the AA Tau 
stellar velocity on the following night.  
We spent 2331 seconds of on-source integration time on AA Tau 
on the night of 30 October 2007, with our optics shifted to enhance 
the sensitivity on the blue side of the 
[NeII] line.

The AA Tau data from the two separate nights were combined by
interpolating onto a common wavelength scale and, in the spectral 
regions where data from
both nights were available, weighting the contribution from 
each night. 
The weighting was used to account for the varying noise 
across the spectra in the overlap region; the variation 
resulted primarily from the rising or falling blaze efficiency.  
We therefore weighted each data point 
inversely as the square of the noise, which was calculated 
assuming that photon noise dominates.  This weighting 
is appropriate if the noise in the two 
spectra being combined add in quadrature. 
In practice, the exact value of the weighting factor had little effect 
on the combined spectrum.

\section{Results}

Figures 1 and 2 show the resulting [NeII] spectra of AA Tau 
and GM Aur. 
The spectra shown have been smoothed by 3 pixels.  Because 
3 pixels is the width of a spectral resolution element at the 
wavelength of [NeII], the smoothing has little impact on our 
results. 
Although the larger noise on the blue side of the [NeII] 
line in AA Tau makes it difficult to characterize the line 
profile with certainty, the emission appears to be  
centered near the stellar heliocentric 
velocity ($v_{\rm helio} = 16.5 \kms$; Bouvier et at.\ 1999) 
and is broad,
extending both redward and blueward of the line center 
with FWHM $\sim 70 \kms$. 
The line appears approximately double-peaked, 
with a blueshifted peak possibly further from the stellar 
velocity and possibly weaker than the redshifted peak.  
The emission redward of the stellar velocity has an equivalent 
width of $6.6\pm 1.4$\,\AA.

In comparison, the spectrally unresolved [NeII] feature detected 
in the $R$=600 {\it Spitzer} IRS spectrum of AA Tau (Carr \& Najita 2008) 
has an equivalent width of approximately 24\,\AA. 
If the emission blueward of the stellar velocity in the TEXES 
spectrum has 
an equivalent width equal to that of the redward  emission, the 
total equivalent width (red and blue) 
is $\sim 0.55$ of 
the equivalent width seen in the {\it Spitzer} spectrum.  
Such a difference might arise if either the mid-infrared 
continuum or the [NeII] emission is time variable. 
Another possibility is that the {\it Spitzer} spectrum 
(4.7\arcsec\ slit width) includes spatially extended [NeII] emission 
that is excluded in the narrower slit (0.6\arcsec) of the TEXES observation. 
Spatially extended [NeII] emission can arise in outflow sources 
(e.g., Neufeld et al.\ 2006; van Boekel et al.\ 2009).  

Spectrally unresolved emission from other lines might also enhance the 
equivalent width of the [NeII] feature in the {\it Spitzer} spectrum. 
However, known emission features are unlikely to contribute 
at this wavelength. 
The synthetic spectrum of AA Tau presented in 
Carr \& Najita (2008) provides a good fit to the molecular 
emission detected in the {\it Spitzer} 
spectrum of that source.  When examined at high spectral resolution, 
the synthetic spectrum reveals no strong lines of 
H$_2$O, HCN, or C$_2$H$_2$ within $200\kms$ of the [NeII] line.  
We therefore interpret the emission features detected with TEXES 
as [NeII]. 

The [NeII] emission from GM Aur has a component of modest width 
(FWHM $\sim 14\kms$) that is centered near the systemic velocity 
determined from measurements of the surrounding molecular disk 
($v_{\rm helio} = 14.8\kms$;  
Dutrey et al.\ 1998; Simon et al.\ 2000).  
There is possible evidence for an additional emission component 
centered $\sim 40\kms$ redward of the stellar velocity,  
although it is difficult to be certain because of the limited 
signal-to-noise ratio of the spectrum. 
The $40\kms$ component might arise from either infalling gas 
located close to the star or gas in rotation about 
the star.  In the latter case, we would expect to 
see an additional emission component with a similar $\sim 40\kms$ 
velocity offset on the blue side of the line. 
Because we observed the [NeII] emission in only one order, 
the GM Aur spectrum has much higher noise on the blue side of 
the line (see \S 2), and it is impossible to determine whether 
a corresponding blueshifted component is present. 
A more complete study of the [NeII] emission from GM Aur 
is needed to distinguish between these possibilities. 

The emission feature centered near the stellar velocity has 
an equivalent width of $18\pm 2$\AA, and the feature 
$\sim 40 \kms$ redward of the stellar velocity has an 
equivalent width of $8\pm 2$\AA.  
In comparison, the equivalent width of the spectrally unresolved 
[NeII] feature detected in an $R$=600 {\it Spitzer} IRS spectrum 
of GM Aur (Najita et al., in preparation) is approximately 70\AA, 
almost three times the combined equivalent width of the two 
[NeII] emission features detected at high spectral resolution. 
As discussed above, variability in the [NeII] line or 
the mid-infrared continuum, alternatively spatially extended 
[NeII] emission, might account for the difference in the 
equivalent widths. 
We are also insensitive to very broad [NeII] emission, 
particulary given the limited signal-to-noise ratio in the 
continuum.  A line 10\% above the marked continuum in Figure~2
and $300\kms$ wide, which is within a {\it Spitzer} resolution element, 
would have an equivalent width of 13\AA, similar to the 
equivalent width reported for the [NeII] line.  Such a broad 
component would be difficult to detect with our data.  

Another possibility is spectrally unresolved emission from other 
lines.  In the {\it Spitzer} spectrum of GM Aur, the emission feature 
detected at the wavelength of [NeII] is broader than an unresolved 
line.  Perhaps the feature includes a contribution from  
lines other than [NeII].  
Mid-infrared molecular features such as H$_2$O, 
HCN, or C$_2$H$_2$ would contribute very negligibly to our 
TEXES observations, because 
any such emission features, if present in the {\it Spitzer} spectrum 
of GM Aur, are much weaker than in AA Tau.
We therefore interpret the emission features detected with TEXES 
as [NeII] for the purpose of this paper.

Thus, as in the case of TW Hya (Herczeg et al.\ 2007), the [NeII]
line profiles of AA Tau and GM Aur are consistent with emission
centered near the stellar velocity.
In comparison with the intrinsic (deconvolved) line width of 
$21\kms$ (FWHM) obtained by Herczeg et al.\ (2007) for TW Hya, 
the width of the [NeII] emission from AA Tau 
is significantly broader ($\sim 70 \kms$).  
While the [NeII] emission from GM Aur has a velocity component centered 
on the stellar velocity that is more similar in width ($\sim 14\kms$) 
to the [NeII] emission from TW Hya, the [NeII] emission may also 
include a redshifted emission component that extends 
to $\gtrsim 50\kms$ of the stellar velocity.

\section{Discussion}

\subsection{Comparison to Profiles of Other Emission Lines}

\subsubsection{H$\alpha$ Profiles}

The double-peaked shape of the [NeII] line profile of AA Tau is 
similar to the shape of the H$\alpha$ line profile of AA Tau 
(Bouvier et al.\ 2007, 2003, 1999; 
Alencar \& Basri 2000; Edwards et al.\ 1994). 
In T Tauri stars, the Balmer lines are believed to form in 
stellar magnetospheres (e.g., Calvet \& Hartmann 1992; Edwards et al.\ 1994; 
Muzerolle et al.\ 1998). 
Although both the H$\alpha$ and [NeII] profiles of AA Tau are 
double-peaked, the H$\alpha$ profile is also much 
broader (FWHM $\sim 280\kms$; Bouvier et al.\ 1999). 

For GM Aur, the H$\alpha$ emission is centrally peaked (Edwards et al.\ 
1994), as is the central component of its [NeII] emission. 
However, the H$\alpha$ emission 
(FWHM of $\sim 220\kms$; Edwards et al.\ 1994) is also much 
broader than the [NeII] emission ($\sim 14\kms$ FWHM); 
the half-width at zero intensity (HWZI) 
of the H$\alpha$ line ($\sim 500\kms$) 
is also much broader than the velocity extent of the redward emission 
of the [NeII] emission ($\sim 50\kms$). 
While the H$\alpha$ and [NeII] profiles have some morphological 
similarities, the difference in the velocity widths suggests that 
the [NeII] emission arises primarily from a different region than 
the stellar magnetosphere.

\subsubsection{CO Fundamental and UV H$_2$ Profiles}

The CO fundamental lines from AA Tau are also double-peaked, 
similar to the shape of the [NeII] line in AA Tau. 
CO fundamental emission is 
believed to probe the inner regions of T Tauri disks 
(see Najita et al.\ 2007 for a review). 
Figure 3 compares the [NeII] profile of AA Tau with 
the average line profile of CO emission from AA Tau that 
was obtained by J. Carr, J. Najita, and N. Crockett
on the night of 26 Nov 2004 using NIRSPEC on the Keck II telescope 
on Mauna Kea.  
The line profile shown is the average of 4 $v$=1--0 lines in the 
$4.9\micron$ region.  Regions of poor telluric correction 
($\lesssim 85$\% transmission, as determined from observations 
of the telluric standard) were excluded from the average.
The double-peaked CO line profile has a FWHM of 
$\sim 140 \kms$ and a peak-to-peak separation of $\sim 80\kms$. 
Thus, the CO profile is broader than the [NeII] line profile, 
but narrower than the Balmer line profile.
A preliminary analysis of the basic properties of the 
CO fundamental emission from AA Tau finds that the emission is 
optically thick and has an average temperature of $\sim 900$\,K  
(Carr \& Najita 2008).  
A more detailed analysis of the CO emission will be presented in a 
future publication. 

CO fundamental emission has been reported from GM Aur by 
Salyk et al.\ (2007), although it was not detected in an earlier 
study (Najita et al.\ 2003). 
Because of the low infrared continuum excess of GM Aur 
compared to that of AA Tau (GM Aur is a transition object), 
the stellar photosphere makes a more significant contribution 
to the GM Aur spectrum than it does in the case of AA Tau.  
We therefore first corrected for stellar photospheric absorption 
in the Salyk et al.\ (2007) spectrum before constructing 
the average CO line profile.  Following the 
approach described in Najita et al.\ (2008), we 
created a synthetic stellar spectrum that is appropriate for 
the stellar spectral type of the source (K3V; Herbig \& Bell 1988). 
We also used a stellar rotational velocity of 
$v\sin i=8\kms $ (cf.\ Hartmann et al.\ 1986). 
We found a good fit to the observed structure in the continuum 
(i.e., in the region away from the 1--0 CO lines) with a veiling 
of 1 at $4.7\micron$.  
This value for the veiling at $4.7\micron$ is consistent with that 
implied by the SED of GM Aur (Furlan et al.\ 2006). 
We subtracted the veiled stellar continuum from spectrum,  
and we then added back an 
equivalent (featureless) continuum in order to show the 
strength of the emission relative to the continuum.  

Figure 4 compares the [NeII] profile of GM Aur with 
the average CO line profile constructed from the resulting spectrum   
in the region of the $v$=1--0 P8 through P12 transitions.  
Regions of poor telluric correction 
($\lesssim 75$\% transmission) were excluded from the average.
The average CO line profile differs from that reported by 
Salyk et al.\ (2007) in that 
there is no significant central dip in the average line profile, 
because the stellar photospheric contribution to the 
line profile has been removed. 
Because the average line profile is now more centrally peaked, 
the FWHM of the average line profile ($\sim 30\kms$) is also smaller than 
the value of $\sim 50 \kms$ reported by Salyk et al.\ (2007). 

The [NeII] and CO line properties can also be compared with those 
of UV fluorescent H$_2$ emission, another line diagnostic that is 
believed to probe the conditions in inner circumstellar disks 
(e.g., Herczeg et al.\ 2006 and references therein; 
see e.g., Najita et al.\ 2007 for a review).  
In the case of TW Hya, Herczeg et al.\ (2007) found that the 
[NeII] emission width ($21\kms$) is broader than 
the UV H$_2$ emission ($14 \kms$), which is broader than 
the CO fundamental emission ($\sim 8\kms$; Salyk et al.\ 2007).  
A similar decreasing sequence is inferred for the nominal 
temperatures of the emitting gas that gives rise to these 
features:  $\sim 4000$\,K for [NeII] (Glassgold et al.\ 2007); 
$\sim 2000$\,K for H$_2$ (Herczeg et al.\ 2004);  
and $\sim 800$\,K for CO (Salyk et al.\ 2007).  
Such a progression might occur if disk atmospheres experience 
transsonic turbulence, which is enhanced in the warmer, atomic 
regions higher up in the disk atmosphere. 
While this interpretation might be considered for a face-on system 
such as TW Hya, the progression of line widths in TW Hya, 
from broader [NeII] lines to narrower H$_2$ and CO lines,  
would not be expected to hold in higher inclination systems 
where disk rotation would play a larger role.

Indeed, for AA Tau, we find that the width of the CO emission is 
instead {\it broader} than the [NeII] emission. 
For GM Aur, the CO line (FWHM $\sim 30\kms$) is also broader 
than the central component of the [NeII] line (FWHM $\sim 14\kms$); 
the HWZI of the average CO line profile could be as large as 
$\sim 50\kms$ (Salyk et al.\ 2007),  
comparable to the velocity extent 
of the red component of the [NeII] emission ($\sim 50\kms$). 
Thus, the CO emission is broader or comparable to the width 
of the [NeII] emission in the two sources studied. 
If both diagnostics are dominated by disk rotation, this suggests that 
the [NeII] emission arises from larger disk radii on average 
than the CO.  

This is consistent, in general, with models of  
the ionization and thermal structure of T Tauri disks (e.g., 
Glassgold et al.\ 2004, 2007; Meijerink et al.\ 2008).  
In the Glassgold et al.\ models, 
the surface of the disk is heated and ionized by stellar X-rays, 
producing a low column density ($N_{\rm H} \sim  10^{20}\persqcm$) 
surface layer of hot ($\sim 4000$\,K)
gas that extends to $\gtrsim 10$\,AU.  Significant [NeII] is 
found to emerge from the disk surface region within $\sim 20$\,AU 
(Glassgold et al.\ 2007; Meijerink et al.\ 2008). In contrast, 
CO becomes abundant at much larger disk vertical column densities 
($N_{\rm H} >10^{20}\persqcm$), and the warm temperatures and high 
densities needed to produce the CO emission ($> 500$\,K) 
likely limits the emission to smaller disk radii, within a 
few AU.

Although for TW Hya the width of the UV H$_2$ emission is 
broader than the CO fundamental emission, 
the inverse is true for other sources with UV H$_2$ and CO 
fundamental emission line widths that have have been reported 
in the literature (see Table 1). 
Thus, as might be expected, the progression of line widths 
found for TW Hya among the [NeII], H$_2$, and CO emission 
lines does not appear to be typical of T Tauri stars.

\subsection{Origin of [NeII] Emission}

Herczeg et al.\ (2007) offered three possible interpretations 
for the origin of the width of the [NeII] line observed from TW Hya: 
(1) Keplerian rotation from the very inner disk region, 
(2) transsonic turbulence in a $\sim10,000$\,K disk atmosphere, 
or (3) photoevaporation at $\sim 10\kms$ from both faces of 
a face-on gaseous disk that is optically thin in the mid-infrared continuum. 
While each of these explanations could, in principle, account for 
the observed $\sim 21 \kms$ line width of TW Hya given its 
low inclination, they would produce different line profiles for 
systems viewed at higher inclinations.  

At higher inclination, a line profile dominated by disk rotation 
would remain symmetric about the stellar velocity but would 
become broader. 
In comparison, a line profile dominated by turbulence would show 
little difference when viewed at higher inclination. 
In the case of a photoevaporative flow 
from a disk that is optically thick in the mid-infrared continuum, 
the disk would occult the receding flow, producing a line profile 
that is blueshifted by up to $\sim 10\kms$, the nominal 
velocity of a photoevaporative flow (e.g., Font et al.\ 2004).  

A physical picture analogous to this last scenario is invoked 
to account for the high velocity 
blueshifted components of the [OI]\,6300\AA\ line emission from 
T Tauri stars 
(e.g., Edwards, Ray, \& Mundt 1993); although 
the high velocities observed in the [OI] case indicate a dynamical 
(rather than thermal) ejection process.  
Font et al.\ (2004) studied whether a lower velocity 
thermally-driven disk photoevaporative flow could account for the 
properties of the low velocity component of the [OI] emission. 
A more recent paper explores the disk photoevaporative flow hypothesis 
specifically for [NeII], in both the case of a continuous disk 
and a disk with an inner hole (Alexander 2008). 
In a system viewed edge-on, the [NeII] line profile is 
predicted to be dominated by Keplerian disk rotation and 
to be centered on the stellar velocity.  At lower inclinations, 
the [NeII] line profile is predicted to be blueshifted from 
the stellar velocity by $5-10\kms$. 

Of the three explanations invoked to explain the width of the 
[NeII] line of TW Hya, only disk rotation could plausibly account 
for the broad profile of the [NeII] emission from AA Tau. 
The width of the emission greatly exceeds the width expected 
from either disk turbulence or photoevaporation. 
The apparent double-peaked line profile might indicate that the 
emission arises from a limited range of disk radii, similar to 
the interpretation given to the double-peaked CO overtone 
lines in sources such as WL16 (Carr et al.\ 1993; Najita et al.\ 1996). 
For Keplerian rotation about a $0.7\Msun$ star viewed at an 
inclination of 75 degrees, 
the redward velocity extent of the [NeII] emission 
and the redward emission peak (to $+45\kms$ and at $\sim 18\kms$ 
from the stellar velocity, respectively) would indicate emission 
extending from an inner radius of $\sim 0.3$\,AU out to 
an outer radius of $\sim 1.8$\,AU.

The narrow ($\sim 14\kms$ FWHM) component of the [NeII] emission 
from GM Aur that is centered at the stellar velocity 
could plausibly be explained by turbulent broadening.  
It might also arise from a rotating Keplerian disk at radii 
beyond several AU 
given the stellar mass ($0.84-1.0\Msun$) and 
inclination ($i=54$ degrees) measured for the system 
(Simon et al.\ 2000; Dutrey et al.\ 2008).  
The lack of a velocity shift of this component from the stellar 
velocity argues against photoevaporation playing a significant 
role in the emission (cf.\ Font et al.\ 2004; Alexander 2008).  
The high velocity of the redshifted component could plausibly 
be accounted for by disk rotation if an (undetected) blueshifted 
component is also present.  The $\sim 40\kms$ velocity shift 
measured for this component would correspond to gas in 
Keplerian rotation at an orbital radius of 0.3\,AU. 

In comparison, Glassgold et al.\ (2007) predicted that an X-ray 
irradiated disk atmosphere would produce [NeII] emission from 
radii within $\sim 20$\,AU.  [NeII] emission is expected to arise from a 
similar range of disk radii in the case of a disk irradiated by 
stellar UV photons (Gorti \& Hollenbach 2008; Alexander 2008). 
In an updated study that extends the Glassgold et al.\ (2007) 
calculation inward to 0.25\,AU, Meijerink et al.\ (2008) 
showed the predicted [NeII] emissivity per unit radius, $P,$ on  
a velocity scale $v_{\rm K}$, where $v_K$ is the Keplerian rotational 
velocity at a given radius. 
This radial contribution function corresponds to a 
[NeII] line profile with a FWHM approximately equal to 
$0.75\,v_1 \sin i$, where 
$v_1$ is the Keplerian velocity at 1\,AU.
The line width is smaller than the width of the 
$P(v_{\rm K})$ distribution because the line profile accounts for 
how the flux from a given annulus, when viewed at an inclination $i$, 
is distributed over a range of velocities between 0 and $v_{\rm K}$. 

For the stellar mass and inclination of GM Aur, the 
width of the [NeII] line is expected to be $17 \kms$, 
similar to the observed $14\kms$ width of the 
central [NeII] emission component of GM Aur. 
Figure 5 compares the observed [NeII] profile of GM Aur 
with a model profile constructed from the 
$P(v_{\rm K})$ distribution of Meijerink et al.\ (2008),  
where the emission extends 
from an outer radius of $\sim 25$\, AU, beyond which the [NeII] 
emission is insignificant, in to the minimum radius 
of 0.25\,AU studied by those authors (short dashed blue line). 
The red-shifted [NeII] emission component extends to higher 
velocities, $\sim 40\kms$, and is not well accounted for 
by the model profile.  
For the $i= 75$ degree inclination of AA Tau, the Meijerink et al.\ (2008) 
results would suggest a 
width for the [NeII] line of $15-18 \kms$ 
for a stellar mass of $0.5-0.7\Msun$.
The observed width of the [NeII] emission from AA Tau is 
significantly broader than this value.

The difference between the observed profiles and the theoretical 
predictions may be due to the simple physical picture 
adopted in the disk emission models; real systems are 
likely to be more complex. 
Glassgold et al.\ (2004) adopted the conventional (flared, 
azimuthally symmetric) disk geometry of D'Alessio et al.\ (1999). 
In contrast, in AA Tau the modulation of its photometric and 
spectroscopic properties suggest that the inner 
disk edge (near the corotation radius) has a large scale 
height over a range in azimuth (Bouvier et al.\ 1999, 2003, 2007).  
Periodic variations in the optical light curve occur in phase 
with enhanced redshifted absorption in the Balmer line profiles.  
These and other observed properties have been interpreted as 
evidence for an optically thick occulting screen produced by 
a magnetically-warped dusty inner disk edge. 
In a system with an 
inflated inner disk edge, more of the stellar ionizing photons 
(X-rays or UV) may be deposited close to the star, resulting 
in [NeII] emission from smaller radii and higher disk 
rotational velocities.  

GM Aur is also an unusual source in that, like TW Hya, it is 
a transition object. 
Based on the shape of the SED, Calvet et al.\ (2005) estimate 
that the disk is devoid of dust in the radial range 5--24\,AU,  
with a small amount of dust present within 5\,AU.
Earlier studies of the SED found a smaller optically thin 
region out to $\sim 4$\,AU (Rice et al.\ 2003; 
see also Bergin et al.\ 2004). 
These authors speculate that the optically thin region 
arises from the presence of a companion that has carved out 
a gap in the disk (see also Marsh \& Mahoney 1992). 

In comparison, the [NeII] emission component centered on the 
systemic velocity in GM Aur has a HWZI of $\sim 10\kms$.  
If the width of the emission arises from Keplerian rotation, 
the emission arises from $\gtrsim 4$\,AU, given the stellar mass 
and inclination of GM Aur (Dutrey et al.\ 1998; Simon et al.\ 2000). 
This component may therefore arise from 
within the optically thin region (5--24\,AU) in the 
interpretation of the SED favored by Calvet et al.\ (2005; 
see also Dutrey et al.\ 2008), or from 
the inner region of the outer disk, if 
the outer disk extends in to the smaller radii favored 
by Rice et al.\ (2003).
Figure 5 shows a model profile constructed using the 
$P(v_{\rm K})$ distribution of Meijerink et al.\ (2008) 
where the emission extends in to 4\,AU (long dashed green line). 
Higher signal-to-noise ratio data are needed to 
determine if this model profile provides a better fit than  
one in which the inner radius extends in to much smaller radii 
(e.g., 0.25\,AU; short dashed blue line). 

The redward component of the [NeII] emission extends to 
$\sim 33-55\kms$ from the stellar velocity.  
If the [NeII] line profile is symmetric across the stellar 
velocity and arises from Keplerian rotation, the redward 
emission component can be inferred to arise from disk radii 
$0.16-0.5$\,AU. 
This component may represent emission from a gaseous inner 
disk close to the star.  Such a gaseous disk is expected to 
be present since GM Aur experiences continued stellar accretion 
at a rate typical of T Tauri stars ($\sim 10^{-8}\Msunperyr$; 
Najita et al.\ 2007).

\section{Summary and Future Directions}

The observations reported here bring to a grand total of four 
the number of T Tauri [NeII] emission sources that have been 
studied at high spectral resolution. 
In the high accretion rate system T Tau, the [NeII] 
emission is associated primarily with outflow activity 
(van Boekel et al.\ 2009). 
In the three lower accretion rate cases studied 
thus far (TW Hya, AA Tau, GM Aur), the 
[NeII] emission appears to be centered at the stellar velocity, 
consistent with a disk origin for the emission. 
The [NeII] emission is broader in the non-face-on systems 
(AA Tau, GM Aur) than in the face-on TW Hya, a result that is 
also consistent with a disk origin for the emission. 

In the non-face-on systems, the [NeII] 
emission width is narrower than (or comparable to) the 
CO emission width.  If the widths of both diagnostics are 
dominated by Keplerian rotation, this suggests that the [NeII] 
emission arises from larger disk radii on average than the 
CO emission.  This is consistent, in general, with models of  
the ionization and thermal structure of T Tauri disks (e.g., 
Glassgold et al.\ 2007; Meijerink et al.\ 2008). 

The observed [NeII] line profiles show little evidence for 
an origin in a disk photoevaporative flow. 
This is particularly so in the case of GM Aur, where 
the central component of the [NeII] emission 
is narrow enough to plausibly arise in a photoevaporative flow, but 
the profile is centered at the stellar velocity rather than 
showing the $5-10\kms$ blueshift 
that is predicted for emission arising in a 
photoevaporative flow (Alexander 2008).  
Photoevaporative flows may be present in these systems 
but contribute a small fraction of the [NeII] emission. 
More recent work by Pascucci \& Sterzik (2009) finds stronger 
evidence for photoevaporative flows in other [NeII]-emitting systems.

The equivalent width of the [NeII] emission we detect is less 
than that of the spectrally unresolved [NeII] feature in the 
{\it Spitzer} spectra of the same sources.  Variability 
in the [NeII] emission or the mid-infrared continuum, 
a spatially extended [NeII] component, or a very (spectrally) broad 
[NeII] component might account for the difference in the 
equivalent widths.  Further work is needed to understand the 
origin of this discrepancy.  

These results illustrate the ability of high resolution 
spectroscopy to probe the origin of the [NeII] emission 
from T Tauri stars. 
Further measurements of resolved [NeII] line profiles, 
for both classical T Tauri stars and transition objects, 
are needed to determine whether the results obtained here 
apply to the majority of [NeII]-emitting T Tauri stars.  
In addition, 
higher sensitivity line profiles 
than those reported here would be useful 
to probe the origin of the [NeII] emission in detail.


%% The \notetoeditor{TEXT} command allows the author to communicate
%% information to the copy editor.  This information will appear as a
%% footnote on the printed copy for the manuscript style file.  Nothing will
%% appear on the printed copy if the preprint or
%% preprint2 style files are used.

%% The eqnarray environment produces multi-line display math. The end of
%% each line is marked with a \\. Lines will be numbered unless the \\
%% is preceded by a \nonumber command.
%% Alignment points are marked by ampersands (&). There should be two
%% ampersands (&) per line.


%% Included in this acknowledgments section are examples of the
%% AASTeX hypertext markup commands. Use \url without the optional [HREF]
%% argument when you want to print the url directly in the text. Otherwise,
%% use either \url or \anchor, with the HREF as the first argument and the
%% text to be printed in the second.

\acknowledgments

We thank Nathan Crockett for his help reducing the CO spectrum of 
AA Tau and Colette Salyk for making available her published CO spectrum 
of GM Aur. 
We thank the Gemini staff for their
support of TEXES observations on Gemini North. The development of
TEXES was supported by grants from the NSF and the NASA/USRA SOFIA
project.  Modification of TEXES for use on Gemini was supported by
Gemini Observatory.  Observations with TEXES were supported by NSF
grant AST-0607312.  
Financial support for the work of JRN and GWD was provided by the 
NASA Origins of Solar Systems program (NNH07AG51I) and 
the NASA Astrobiology Institute under Cooperative Agreement 
No.\ CAN-02-OSS-02 issued through the Office of Space Science.
This work was also supported by the Life and Planets Astrobiology 
Center (LAPLACE). 
Basic research in infrared astronomy at the Naval Research Laboratory
is supported by 6.1 base funding. 
MJR acknowledges support from NSF grant AST-0708074 and NASA grant NNG04GG92G.  
This work is based
on observations obtained at the Gemini Observatory, which is operated
by the Association of Universities for Research in Astronomy, Inc.,
under a cooperative agreement with the NSF on behalf of the Gemini
partnership: the National Science Foundation (United States), the
Particle Physics and Astronomy Research Council (United Kingdom),
the National Research Council (Canada), CONICYT (Chile), the
Australian Research Council (Australia), CNPq (Brazil) and CONICET
(Argentina).
The NIRSPEC data presented herein were obtained at the W.\ M.\ 
Keck Observatory, in part via the TSIP program administered by NOAO 
and in part from 
telescope time allocated to NASA through the agency's scientific 
partnership with the California Institute of Technilogy and the 
University of California.  The Observatory was made possible by the 
generous financial support of the W.\ M.\ Keck Foundation.
The authors wish to recognize and acknowledge the very significant
cultural role and reverence that the summit of Mauna Kea has always
had within the indigenous Hawaiian community.  We are most fortunate
to have the opportunity to conduct observations from this mountain.


%% in parentheses, after the keyword, but they will not be verified.

{\it Facilities:} \facility{Gemini (TEXES)}, \facility{Keck (NIRSPEC)}.

\begin{figure}
\figurenum{1}
\plotfiddle{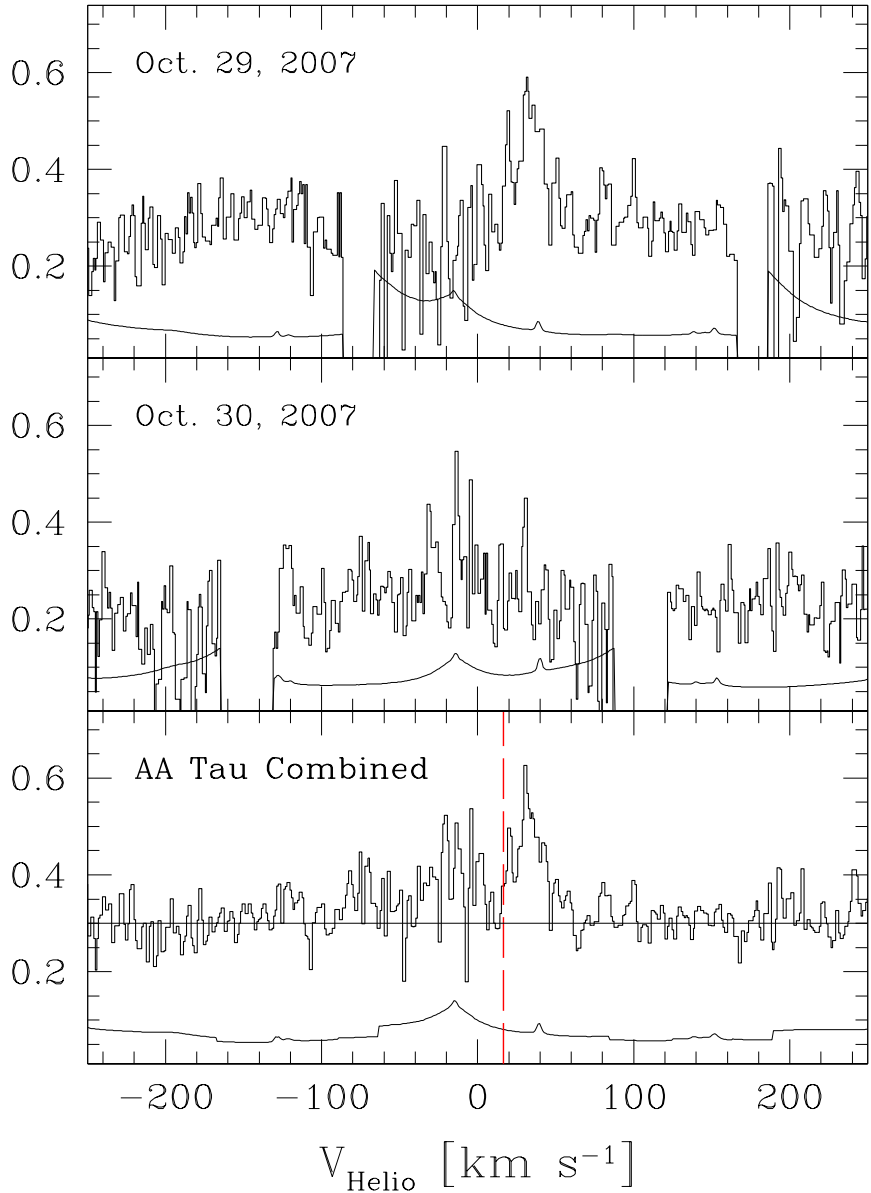}{5.5truein}{0}{80}{80}{-250}{-110}
\caption{Spectrum of [NeII] emission from AA Tau (histograms) 
plotted as a function of heliocentric velocity and smoothed by 3 pixels. 
The vertical scale is flux density in units of Jy.
Spectra obtained on 29 Oct 2007 and 30 Oct 2007 (top and middle panels) 
are shown along with the combined spectrum (bottom panel).  
The light lines in the upper two panels indicate the noise per pixel 
calculated 
by the TEXES pipeline at each point in the spectrum.
The gaps in the spectra in the upper two panels reflect the gaps between 
the TEXES orders at this wavelength.
The vertical dashed line in the bottom panel indicates the stellar velocity. 
}
\end{figure}

\begin{figure}
\figurenum{2}
\plotfiddle{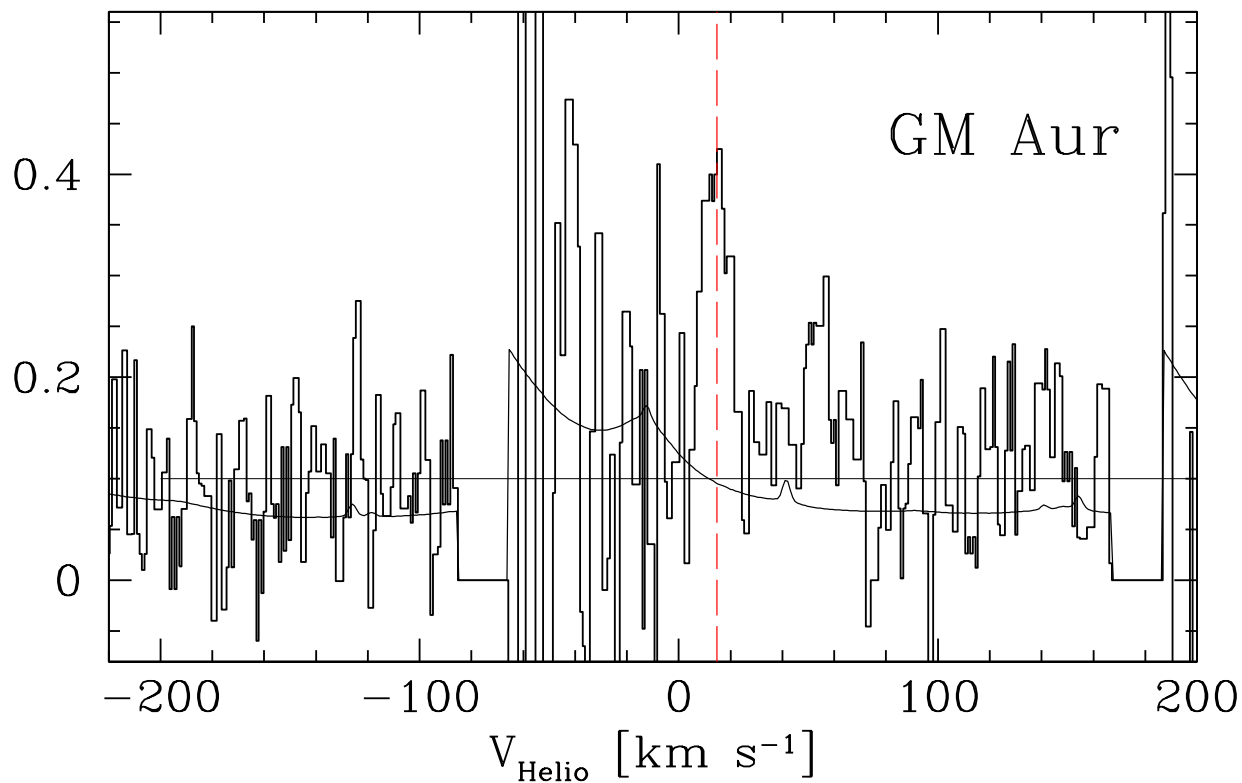}{5.5truein}{270}{70}{70}{-275}{450}
\caption{Spectrum of [NeII] emission from GM Aur 
plotted as a function of heliocentric velocity and smoothed by 3 pixels. 
The vertical scale is flux density in units of Jy.
The vertical dashed line indicates the stellar velocity, 
and the horizontal line indicates the continuum level. 
The light lines indicate the noise per pixel calculated 
by the TEXES pipeline at each point in the spectrum.
In addition to the [NeII] emission component centered on 
the stellar velocity, another emission component appears 
to be present at $+40\kms$ from the stellar velocity. 
}
\end{figure}

\begin{figure}
\figurenum{3}
\plotfiddle{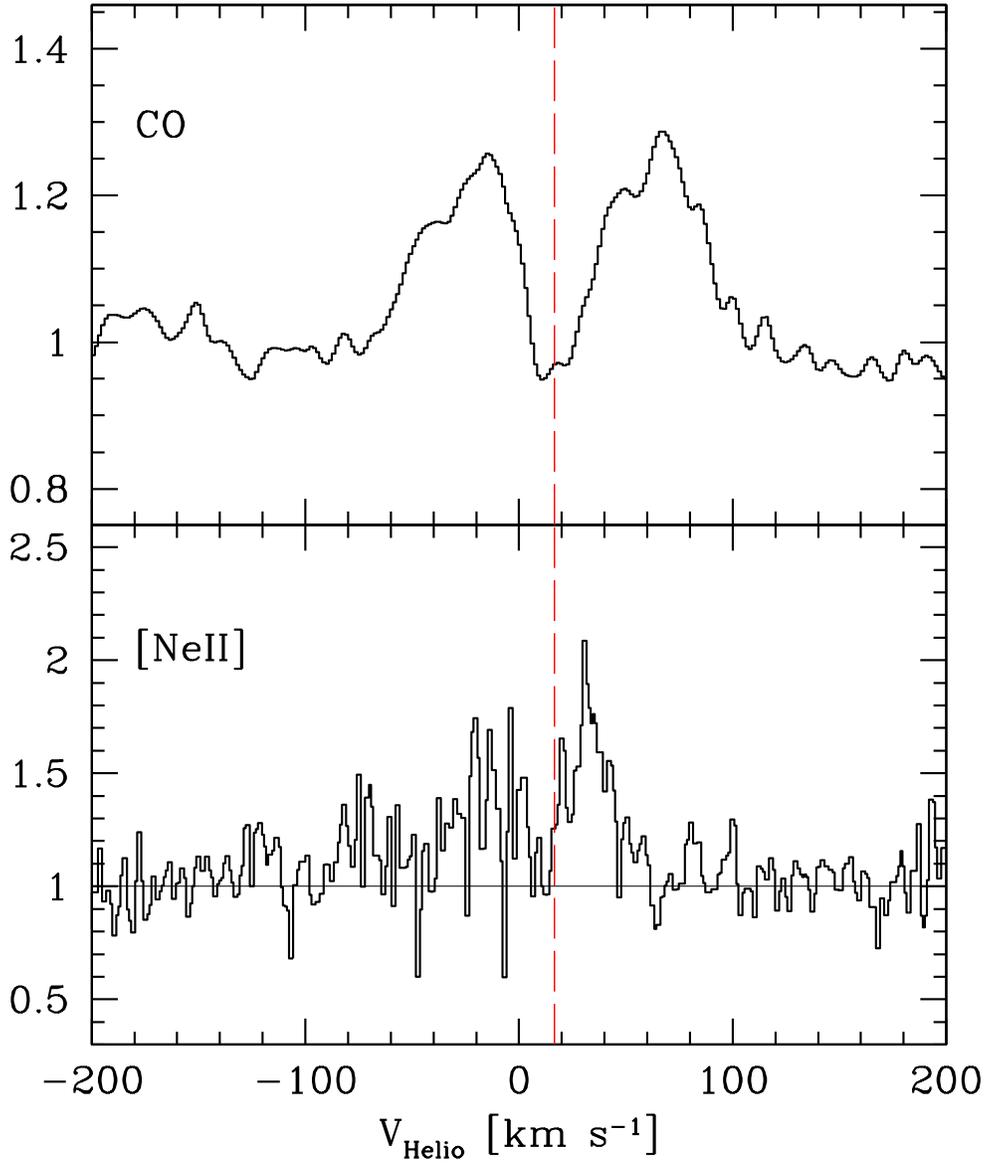}{4.5truein}{0}{80}{80}{-240}{-130}
\caption{{\it Top:} CO emission profile from AA Tau in the 
$4.9\micron$ region, normalized to the continuum.  
The vertical dashed line indicates 
the stellar velocity ($v_{\rm helio} = 16.5\kms$).  
The line profile shown is the average of the 
$v$=1--0 P25 through P28 lines.  Regions 
of poor telluric correction were excluded from the average. 
{\it Bottom:} [NeII] line profile of AA Tau from the combined spectrum shown 
in Figure 1, normalized to the continuum. 
}
\end{figure}

\begin{figure}
\figurenum{4}
\plotfiddle{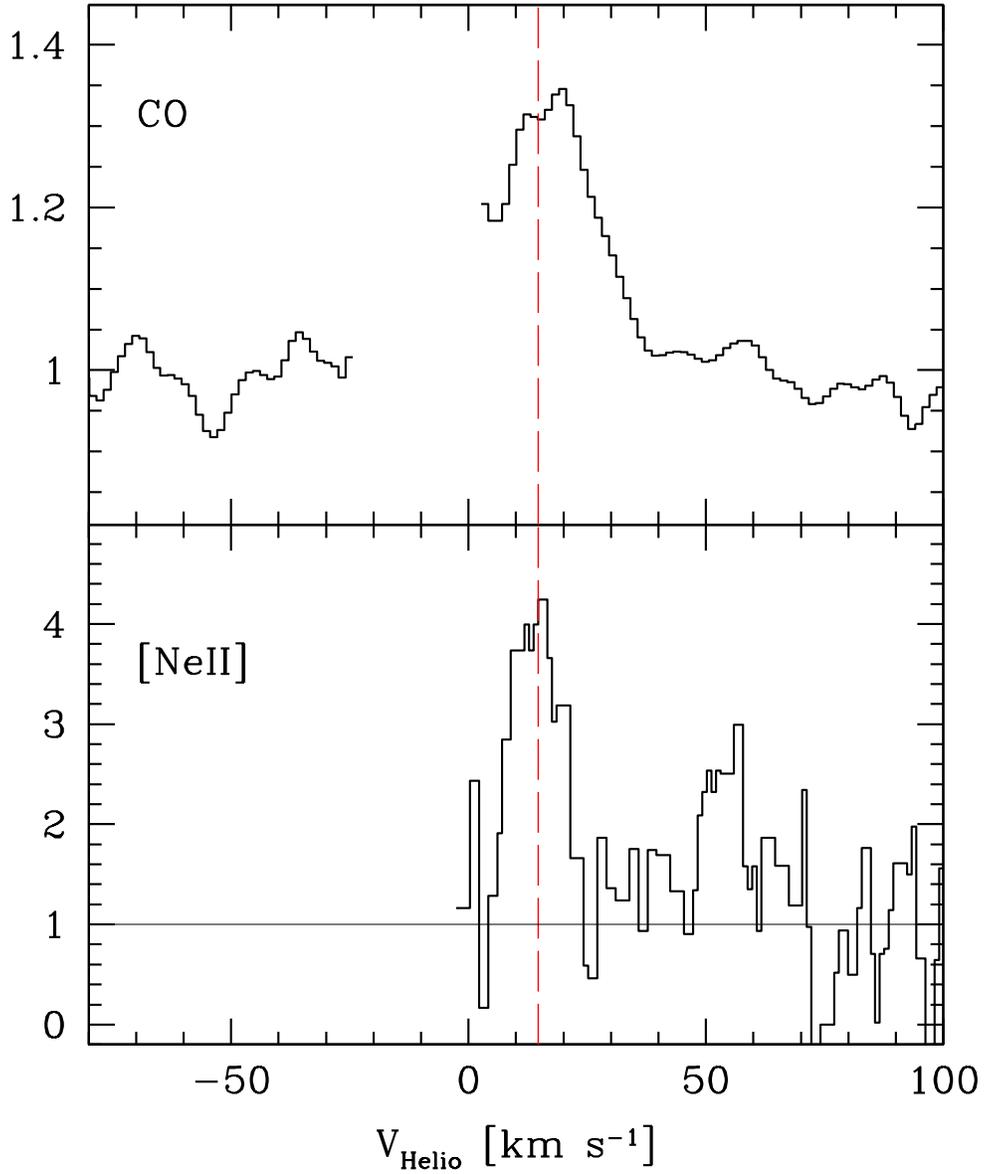}{4.5truein}{0}{80}{80}{-240}{-130}
\caption{{\it Top:} CO emission profile from GM Aur in the 
$4.7\micron$ region, normalized to the continuum.  
The vertical dashed line indicates 
the stellar velocity ($v_{\rm helio} = 14.7\kms$).  
The line profile shown is the average of the 
$v$=1--0 P8 through P12 lines.  In constructing the profile,  
the stellar photospheric contribution to the spectrum was first removed 
and regions of poor telluric correction were excluded from the average. 
{\it Bottom:} [NeII] line profile of GM Aur from the spectrum shown 
in Figure 2, normalized to the continuum. 
}
\end{figure}

\begin{figure}
\figurenum{5}
\plotfiddle{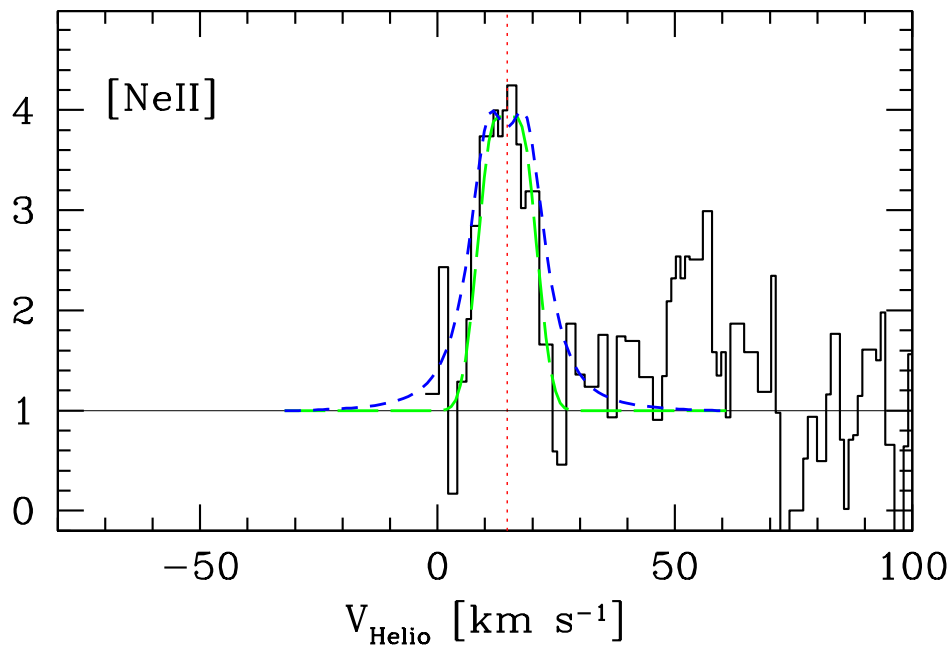}{4.5truein}{0}{80}{80}{-240}{-130}
\caption{[NeII] emission profile of GM Aur, compared with synthetic 
[NeII] disk emission profiles from an X-ray irradiated disk 
(Meijerink et al.\ 2008; see text for details). 
The model profiles assume Keplerian rotation and the stellar mass ($0.85\Msun$)
and inclination ($i=54$ degree) of GM Aur reported by Simon et al.\ (2000). 
The emission extends from an outer radius of $\sim 25$\,AU, beyond 
which the [NeII] emission is insignificant, inward to 
0.25\,AU (short dashed blue line) or 4\,AU (long dashed green line).  
A vertical scaling factor is applied to each model to facilitate comparison 
with the observed profile. 
}
\end{figure}

\begin{deluxetable}{lrcccl} 
\tablecolumns{6} 
\tablewidth{0pc} 
\tablecaption{[NeII], H$_2$, and CO Line FWHM} 
\tablehead{ 
\colhead{Source}  & 
\colhead{$i$}   & 
\colhead{$\sigma$(NeII)} & 
\colhead{$\sigma$(H$_2$)} & 
\colhead{$\sigma$(CO)} & \colhead{Refs.}\tablenotemark{a} \\
\colhead{}        & 
\colhead{}      & 
\colhead{$\kms$}  & 
\colhead{$\kms$}  & 
\colhead{$\kms$}  & \colhead{}
}
\startdata 
AA Tau & 75	& 70	    & \nodata & 145	& B99, Here \\
GM Aur & 54	& 14	    & \nodata & 30 	& S00, S07, Here \\
TW Hya & 4      & 21  	    & 14      & 8	& P08, H07, S07 \\
BP Tau & $<50$	& \nodata   & 57      & 70	& A02, N03 \\
DF Tau & 80	& \nodata   & 27      & 65	& A02, N03 \\
       &   	&           & 23      &        	& H06    \\
RW Aur & 40	& \nodata   & 52      & 80--250	& A02, N03 \\
V836 Tau & 65	& \nodata   & 24      & 130 	& H06, N08 \\
\enddata 
\tablenotetext{a}{A02=Ardila et al.\ 2002; 
B99=Bouvier et al.\ 1999;
H06=Herczeg et al.\ 2006; H07=Herczeg et al.\ 2007; 
N03=Najita et al.\ 2003; N08=Najita et al.\ 2008; 
P08=Pontoppidan et al.\ 2008;
S00=Simon et al.\ 2000; S07=Salyk et al.\ 2007}
\end{deluxetable}

\end{document}